\documentclass[prc,superscriptaddress,unsortedaddress,twocolumn,showpacs,preprintnumbers,amsmath,amssymb,floatfix,dvipdfmx]{revtex4}
\usepackage{amsmath}
\usepackage{amssymb}
\usepackage{times}
\usepackage{mathrsfs}
\usepackage{multirow}
\usepackage{bm}
\usepackage{wrapft}
\usepackage{color}
\usepackage[dvipdfmx]{graphicx}

\def\la{\mathrel{\mathpalette\fun <}}

\def\fun#1#2{\lower3.6pt\vbox{\baselineskip0pt\lineskip.9pt
\ialign{$\mathsurround=0pt#1\hfil##\hfil$\crcr#2\crcr\sim\crcr}}}

\newcommand{\beq}{\begin{equation}}
\newcommand{\eeq}{\end{equation}}
\newcommand{\bea}{\begin{eqnarray}}
\newcommand{\eea}{\end{eqnarray}}

\newcommand{\bfi}[1]{\mbox{\boldmath $#1$}}
\newcommand{\bfis}[1]{\mbox{\boldmath ${\scriptstyle #1}$}}

\newcommand{\vk}{{\bfi k}}

\newcommand{\vik}{{\bfis k}}

\begin{document}

\title{
Effects of a chiral three-nucleon force on nucleus-nucleus scattering
}

\author{Kosho Minomo}
\email[]{minomo@rcnp.osaka-u.ac.jp}
\affiliation{Research Center for Nuclear Physics, Osaka University, Ibaraki 567-0047, Japan}

\author{Masakazu Toyokawa}
%\email[]{toyokawa@phys.kyushu-u.ac.jp}
\affiliation{Department of Physics, Kyushu University, Fukuoka 812-8581, Japan}

\author{Michio Kohno}
%\email[]{kohno@kyu-dent.ac.jp}
\affiliation{Physics Division, Kyushu Dental University, Kitakyushu 803-8580, Japan}
\affiliation{Research Center for Nuclear Physics, Osaka University, Ibaraki 567-0047, Japan}

\author{Masanobu Yahiro}
%\email[]{yahiro@phys.kyushu-u.ac.jp}
\affiliation{Department of Physics, Kyushu University, Fukuoka 812-8581, Japan}

\date{\today}

\begin{abstract} 
We investigate the effects of chiral NNLO three-nucleon force (3NF) on 
nucleus-nucleus elastic scattering, using a standard prescription
based on the Brueckner-Hartree-Fock method and the $g$-matrix folding model. 
The $g$-matrix calculated in nuclear matter from 
the chiral N$^{3}$LO two-nucleon forces (2NF) is close to that from the Bonn-B 2NF.
Because the Melbourne group have already developed a practical $g$-matrix interaction
by localizing the nonlocal $g$-matrix calculated from the Bonn-B 2NF,
we consider the effects of chiral 3NF, in this first attempt to study
the 3NF effects, by modifying the local Melbourne $g$-matrix according to
the difference between the $g$-matrices of the chiral 2NF and 2NF+3NF.
For nucleus-nucleus elastic scattering, the 3NF corrections make the folding
potential less attractive and more absorptive.
The latter novel effect is due to the enhanced tensor correlations in triplet channels.
These changes reduce the differential cross section at the middle and large angles,
improving the agreement with the experimental data for 
$^{16}$O-$^{16}$O scattering at 70 MeV/nucleon and
$^{12}$C-$^{12}$C scattering at 85 MeV/nucleon.
\end{abstract}

\pacs{21.30.Fe, 24.10.Ht, 25.70.Bc}
%21.30.Fe  Forces in hadronic systems and effective interactions
%24.10.Ht  Optical and diffraction models
%25.70.Bc  Elastic and quasielastic scattering

\maketitle

%Introduction
%\section{Introduction}
{\it Introduction.} 
How important are three-nucleon forces (3NFs) in nuclear physics?
This is one of the most important issues.
This long-standing subject started with 
the 2$\pi$-exchange 3NF proposed by Fujita and Miyazawa~\cite{Fuj57}.
For light nuclei, attractive 3NFs were introduced to reproduce 
the binding energies~\cite{Wir02}.
In symmetric nuclear matter, meanwhile, 
repulsive 3NFs were introduced to explain empirical saturation properties; 
for example see Refs.~\cite{Wir88,Mut00,Dew03,Bog05}.

It is quite hard to determine 2NF and 3NF phenomenologically. 
This problem can be solved by chiral effective field 
theory (Ch-EFT)~\cite{Epe05,Epe09,Heb11,mac11}, 
since the theory systematically organizes interactions 
among nucleons. The roles of chiral 3NF was recently analyzed 
in light nuclei~\cite{Epe02,Nog06,Nav07,Ski11}. 
For symmetric nuclear matter, the $g$-matrix calculated 
from chiral N$^{3}$LO 2NF plus NNLO 3NF~\cite{Epe05,Heb11}
well explain the empirical saturation properties 
with no adjustable parameter~\cite{Sam12,Koh12,Car13,Koh13}. 
Scattering and reaction phenomena are 
important processes to explore 3NF effects.

Microscopic understanding of nucleon-nucleus and nucleus-nucleus scattering is 
another important subject in nuclear physics. 
The standard method of describing the scattering is 
the $g$-matrix folding model~\cite{Jeu77,Ber77,Bri77,Yam83,Amo00,Kho00}
with the local-density approximation. 
In the framework, the 3NF effects appear through
the density dependence of the $g$-matrix.
The CEG07 $g$-matrix is constructed from the 2NF based on 
the extended soft core model~\cite{Rij06a,Rij06b} and 3NF effects  
are empirically taken into account by introducing the density-dependent 
vector-meson mass that reproduces empirical saturation properties of 
symmetric nuclear matter~\cite{Fur08,Fur09a,Fur09b}. 
In the $g$-matrix calculation of Ref.~\cite{Raf13}, meanwhile, 
phenomenological 3NFs such as Urbana IX~\cite{Pud97} and 
the three-nucleon interaction model~\cite{Fri81,Lag81} are added 
to AV18 2NF~\cite{Wir95}. 
Since these phenomenological 3NFs are also introduced to reproduce 
saturation properties of symmetric nuclear matter, they make 
the $g$-matrix less attractive and less absorptive 
at densities ($\rho$) higher than the normal density $\rho_0$.
The 3NF effects then reduce the folding potential $U(R)$ 
particularly at small distances ($R$). 
For nucleus-nucleus elastic scattering, the 3NF effects are remarkable 
particularly at the middle and large angles where 
projectile and target densities are largely overlapped, 
although the strength of the imaginary part of $U(R)$ is adjusted 
to reproduce the experimental data in the analyses~\cite{Fur09b}. 
The question to be addressed now is how the effects of chiral 3NF 
appear in nucleus-nucleus elastic scattering.

In this Rapid Communication, we investigate the effects of 
chiral NNLO 3NF on nucleus-nucleus scattering, using 
the Brueckner-Hartree-Fock (BHF) method for nuclear matter and 
the $g$-matrix folding model for the scattering. 
In the nuclear matter calculations, furthermore, 3NFs are treated 
with the mean field approximation~\cite{Loi71,Kas74,Fri81,Hol10}, 
i.e., by averaging 3NFs over the third nucleon in the Fermi sea, 
since it is quite hard to treat three-body correlations in nuclear matter. 
This approximation is considered to be good 
for the low transferred-momentum components of the $g$-matrix. 
We then present the following simple prescription, 
as the first estimate of chiral-3NF effects on nucleus-nucleus scattering. 

For nuclear matter, the single-particle potential ${\cal U}$ is determined from 
the diagonal component of the $g$-matrix. 
The ${\cal U}$ calculated from chiral 2NF is close to that 
from Bonn-B 2NF~\cite{Mac87}. Hence, the effects 
of chiral 3NF can be expressed with the ratio of 
the ${\cal U}$ calculated from chiral 2NF+3NF to that from Bonn-B 2NF. 
For finite nuclei, the Melbourne group already presents a practical 
$g$-matrix interaction by localizing the nonlocal $g$-matrix calculated 
from Bonn-B 2NF~\cite{Amo00}. 
The Melbourne $g$-matrix is highly reliable 
for nucleon-nucleus scattering where the 3NF effects are considered to be 
small~\cite{Raf13}, since 
the folding potential with the Melbourne $g$-matrix well reproduces 
the experimental data with no adjustable parameter~\cite{Amo00,Min10,Toy13}. 
The effects of chiral 3NF are then 
introduced to the local Melbourne $g$-matrix by multiplying it by the ratio.  
With the simple prescription, we make a qualitative discussion about 
the effects of chiral 3NF on nucleus-nucleus scattering. 
We consider $^{16}$O-$^{16}$O scattering at 70 MeV/nucleon
and $^{12}$C-$^{12}$C scattering at 85 MeV/nucleon,
since chiral EFT is good at lower incident energies ($E$).

We first investigate the effects of chiral NNLO 3NF 
on symmetric nuclear matter 
through the $\cal U$ that corresponds to the folding potential $U(R)$ 
in nucleus-nucleus elastic scattering, and 
calculate $U(R)$ from the Melbourne interaction 
with chiral-3NF corrections to analyze the effects 
on nucleus-nucleus scattering.

%\section{BHF calculations}
{\it BHF calculations.} 
Following Ref.~\cite{Koh13}, we evaluate the single-particle potential 
with the BHF method for positive energy $E$ 
from chiral N$^{3}$LO 2NF plus NNLO 3NF with
the cut-off energy $\Lambda=550$ MeV~\cite{Epe05,Heb11}. 
Here the chiral NNLO 3NF $V_{123}$ is reduced to an effective 2NF $V_{12(3)}$ 
with the mean-field approximation, that is, 
by integrating the 3NF over the third nucleon, and 
nuclear matter calculations are done from the combination of 
N$^{3}$LO 2NF $V_{12}$ and $V_{12(3)}$ in the standard manner.

The single-particle potential ${\cal U}^{ST}(k_{\rm F},E)$ thus obtained 
depends on $E$, total spin ($S$) and total isospin ($T$) of two nucleons and 
the Fermi momentum $\hbar k_{\rm F}$. 
The potential is related to the $g$-matrix $G^{ST}$ as 
\bea
{\cal U}^{ST}(k_{\rm F}^{},E)&=&\sum_{\vik'}^{k_{\rm F}^{}}
\langle \vk \vk'| G^{ST} 
\notag \\
&& {} + \frac{1}{6} V_{12(3)}^{ST}(1+\frac{Q}{E-H} G^{ST}) |\vk\vk'\rangle_{A},
\label{spp}
\eea
where $\hbar \vk$ and $\hbar \vk'$ are momenta of correlated two nucleons,
the subscript $|~~\rangle_{\cal A}^{}$ means the antisymmetrization 
between the correlated nucleons, and $Q/(E-H)$ is the nucleon propagator including
the Pauli exclusion operator $Q$; see Ref.~\cite{Koh13} for detail.
The momentum $\hbar \vk$ is related to $E$ 
as $E =(\hbar\vk)^2/(2m) + {\rm Re}~{\cal U}$ for the nucleon mass $m$. 
The potential ${\cal U}^{ST}$ is mainly determined 
from the first term $G^{ST}$ on the right hand side of Eq.~\eqref{spp}.
For example at the normal density, 
the contribution of the first term is ten times as large as 
that of the second term having $V_{12(3)}$. 
In the first term, only the central part of the on-shell component of 
$G^{ST}$ contributes to the ${\cal U}^{ST}$, but the tensor part of 
$V_{12}$ and $V_{12(3)}$ affects the central part significantly. 
The mean-field approximation taken is good 
for the ${\cal U}^{ST}$ that is mainly determined from 
the on-shell component of $G^{ST}$.

\begin{figure}[tbp]
\begin{center}
 \includegraphics[width=0.48\textwidth,clip]{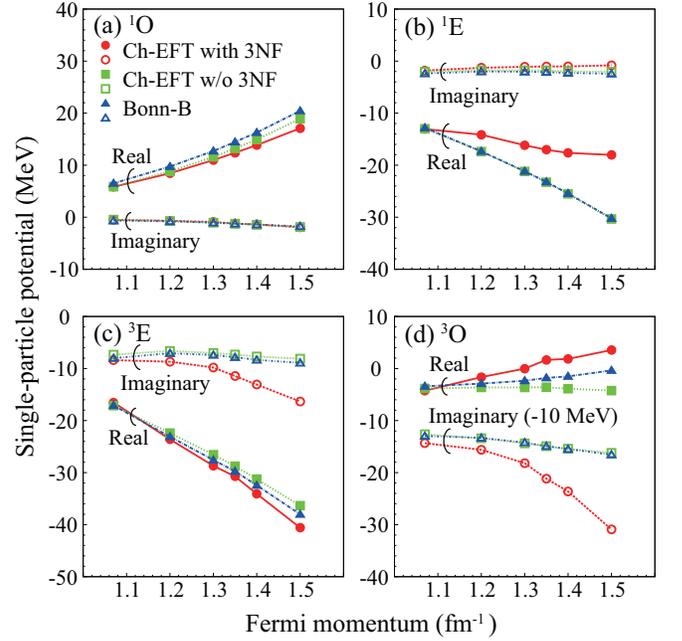}
 \caption{(Color online)
$k_{\rm F}^{}$ dependence of the single-particle potentials at $E=70$ MeV 
for (a) $^1$O, (b) $^1$E, (c) $^3$E, and (d) $^3$O channels.
The circles, squares, and triangles stand for the results of chiral 2NF+3NF, 
chiral 2NF and Bonn-B 2NF, respectively.
The closed (open) symbols correspond to the real (imaginary) part 
of the single-particle potentials.
For $^3$O, the imaginary part is shifted down by 10 MeV. 
}
 \label{fig1}
\end{center}
\end{figure}

% Results of BHF calculations
%\section{Results of BHF calculations}
{\it Results of BHF calculations.} 
Figure~\ref{fig1} displays $k_{\rm F}^{}$ dependence 
of ${\cal U}^{ST}(k_{\rm F},E)$ at $E=70$ MeV for each $S$-$T$ channel.
The results of chiral 2NF (squares) are close to those of 
Bonn-B 2NF (triangles) for all the channels except the $^3$O channel. 
For the real part of ${\cal U}^{ST}$ in the $^3$O channel,   
the difference between the two results is not small particularly 
at high densities, but it little affects angular distributions 
of elastic cross sections, since the magnitude itself is small there.

%----------------------
% Figure 1
%----------------------
%----------------------

Comparing the results of chiral 2NF+3NF (circles) with those of chiral 2NF 
(squares), one can see appreciable changes for the real part of 
${\cal U}^{ST}$ in the $^1$E channel 
and for the real and imaginary parts in the triplet channels ($^3$E and $^3$O).
Chiral 3NF thus makes ${\cal U}^{ST}$ much less attractive for $^1$E.
This effect has been known to be caused
by the suppression of transitions to $\Delta$ resonance due to Pauli blocking. 
For the triplet channels, the 3NF enhances tensor correlations, and thereby 
makes  ${\cal U}^{ST}$ much more attractive and absorptive for $^3$E 
and more absorptive but less attractive for $^3$O. 
As the net effect of all the channels, 
the ${\cal U}^{ST}$ becomes much more absorptive 
because of tensor correlations enhanced by chiral 3NF, and 
becomes less attractive 
because of the suppression of transitions to $\Delta$ resonance in $^1$E. 
The enhancement of tensor correlations due to chiral 3NF is thus 
essential in addition to the repulsive nature of chiral 3NF itself. 
In the region $k_{\rm F} \la 1.1~{\rm fm}^{-1}$ corresponding 
to $\rho \la \rho_0/2$, 
chiral 3NF yields little effect. 
Such 3NF effects on the ${\cal U}^{ST}$ directly reflect nucleus-nucleus 
scattering through the folding potential $U(R)$, as shown latter. 
This is natural, because ${\cal U}^{ST}$ in nuclear matter plays 
the same role as $U(R)$ in nucleus-nucleus scattering.

The Melbourne group constructed the local $g$-matrix interaction $g^{ST}(s;k_{\rm F}^{},E)$ 
from Bonn-B 2NF and showed that the interaction is quite practical and reliable 
for nucleon-nucleus scattering \cite{Amo00}, where $s$ is the coordinate 
between correlated nucleons. 
We then introduce 3NF corrections to $g^{ST}(s;k_{\rm F}^{},E)$ in the following manner, 
since the ${\cal U}^{ST}$ calculated from Bonn-B 2NF well simulate the results 
of chiral N${^3}$LO 2NF:
\bea
g^{ST}(s;k_{\rm F}^{},E)
\rightarrow
f_{}^{ST}(k_{\rm F}^{},E)g^{ST}(s;k_{\rm F}^{},E) 
\label{gst}
\eea
with the factor $f_{}^{ST}(k_{\rm F}^{},E)$ defined by 
\bea
f_{}^{ST}(k_{\rm F}^{},E)=
{\cal U}_{({\rm 2NF+3NF})}^{ST}(k_{\rm F}^{},E)
/{\cal U}_{({\rm 2NF})}^{ST}(k_{\rm F}^{},E) , 
\label{fkf}
\eea
where ${\cal U}_{({\rm 2NF+3NF})}^{ST}(k_{\rm F}^{},E)$ and 
${\cal U}_{({\rm 2NF})}^{ST}(k_{\rm F}^{},E)$ stand for 
the single-particle potentials with and without 3NF, respectively.
This prescription mainly takes account of the effects of 
the on-shell component of chiral 3NF 
on the local Melbourne $g$-matrix interaction.

%\section{Folding-model calculations}
{\it Folding-model calculations.} 
Now we consider nucleus-nucleus scattering with the folding model. 
In the model, the folding potential $U(R)$ is obtained by
folding $g^{ST}(s;k_{\rm F},E)$ with projectile and target 
densities in the standard way~\cite{Sin75,Sin79,Sat79,Kho00,Fur09b}, 
where the Fermi momentum $\hbar k_{\rm F}$ is replaced by a local Fermi momentum 
that is evaluated with the frozen density approximation.
That is, $k_{\rm F}$ is estimated from the sum of 
projectile and target local densities
at the middle point of interacting nucleons.
As the projectile and target proton densities, we take 
the phenomenological ones determined from the electron 
scattering~\cite{Vri87}, applying proton finite-size corrections to 
the densities with the standard manner~\cite{Sin78}. 
The neutron densities are assumed to be the same as the proton ones,
which is accurate enough for light nuclei.

\begin{figure}[tbp]
\begin{center}
 \includegraphics[width=0.35\textwidth,clip]{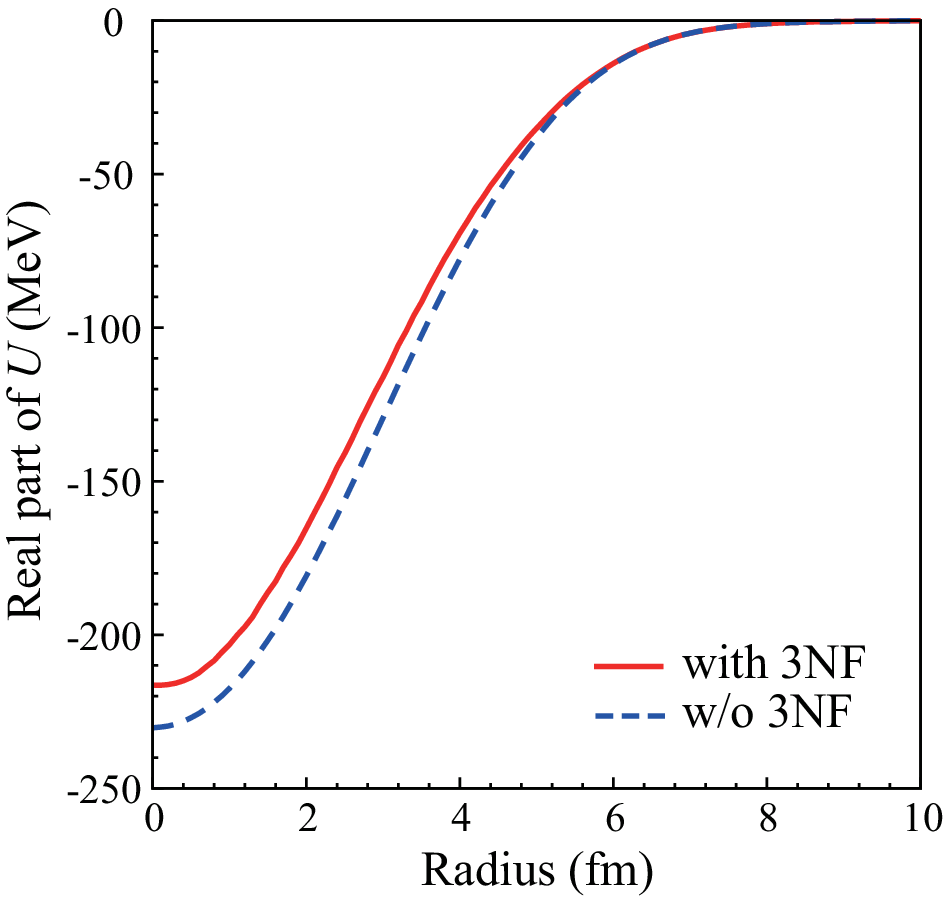}
 \includegraphics[width=0.35\textwidth,clip]{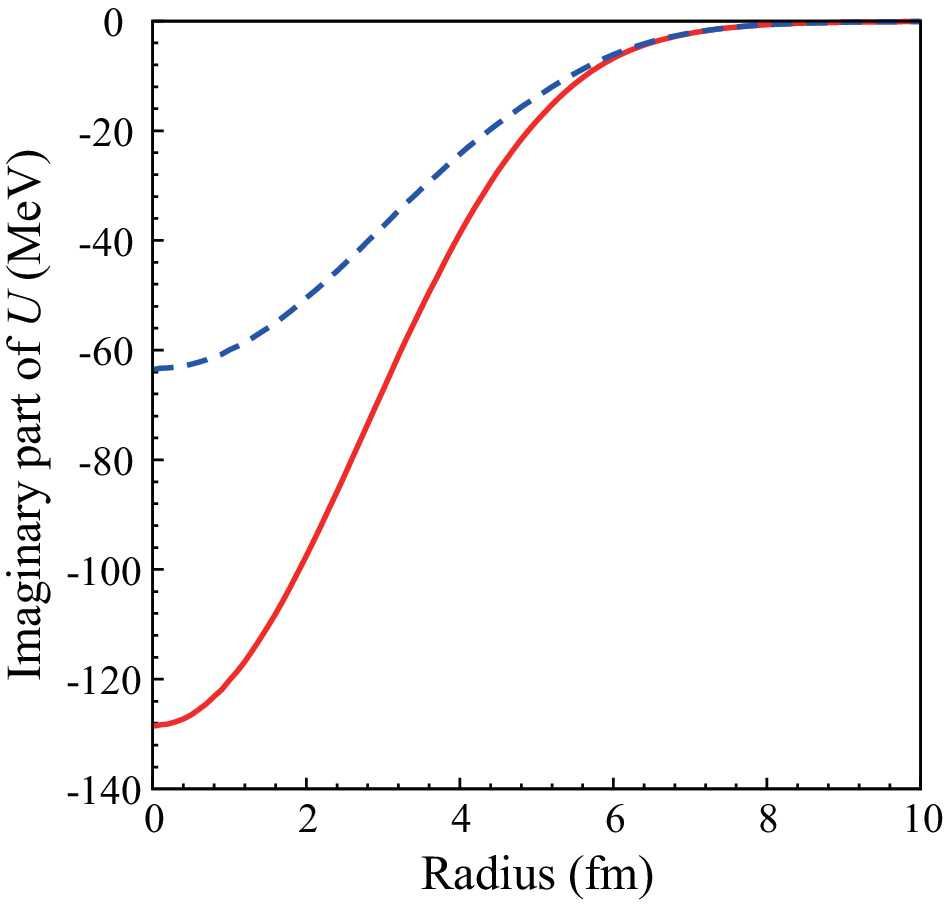}
 \caption{(Color online) 
The folding potential for $^{16}$O-$^{16}$O elastic scattering 
at 70 MeV/nucleon. The real and imaginary parts of the potential are shown 
in the upper and lower panels, respectively. 
The solid (dashed) curve represents the result with (without) 3NF effects.
}
 \label{fig2}
\end{center}
\end{figure}

%\section{Results of folding-model calculations}
{\it Results of folding-model calculations.} 
Figure~\ref{fig2} shows the folding potential $U(R)$
for $^{16}$O-$^{16}$O elastic scattering at 70 MeV/nucleon. 
The solid (dashed) curve represents the result with (without) 3NF corrections.
The potential is mainly determined from the even-channel ($^1$E and $^3$E) 
components, since the odd-channel ($^1$O and $^3$O) components are almost 
canceled after the sum of the direct and exchange parts of $U(R)$.
For the real part of $U(R)$, chiral 3NF works as a repulsive force, 
but it is weak because of the offset between the repulsive contribution in 
$^1$E and the attractive one in $^3$E. 
For the imaginary part, chiral 3NF makes $U(R)$ strongly absorptive, since 
the contribution works additively between $^3$E and $^3$O channels.

%----------------------
% Figure 2
%----------------------
%----------------------

In the previous works of Refs.~\cite{Fur08,Fur09a,Fur09b,Raf13}, 
the phenomenological 3NFs work repulsively and less absorptively. 
For the real part of $U(R)$, the present result is consistent with 
the previous ones qualitatively, although the repulsive effect is small 
in the present work but large in the previous works. 
For the imaginary part, meanwhile, chiral 3NF 
enhances the absorption of $U(R)$,
in contrast to the opposite effect of phenomenological 3NFs. 
The difference comes from the fact that chiral 3NF enhances tensor 
correlations but the phenomenological 3NFs do not.

\begin{figure}[tbp]
\begin{center}
 \includegraphics[width=0.335\textwidth,clip]{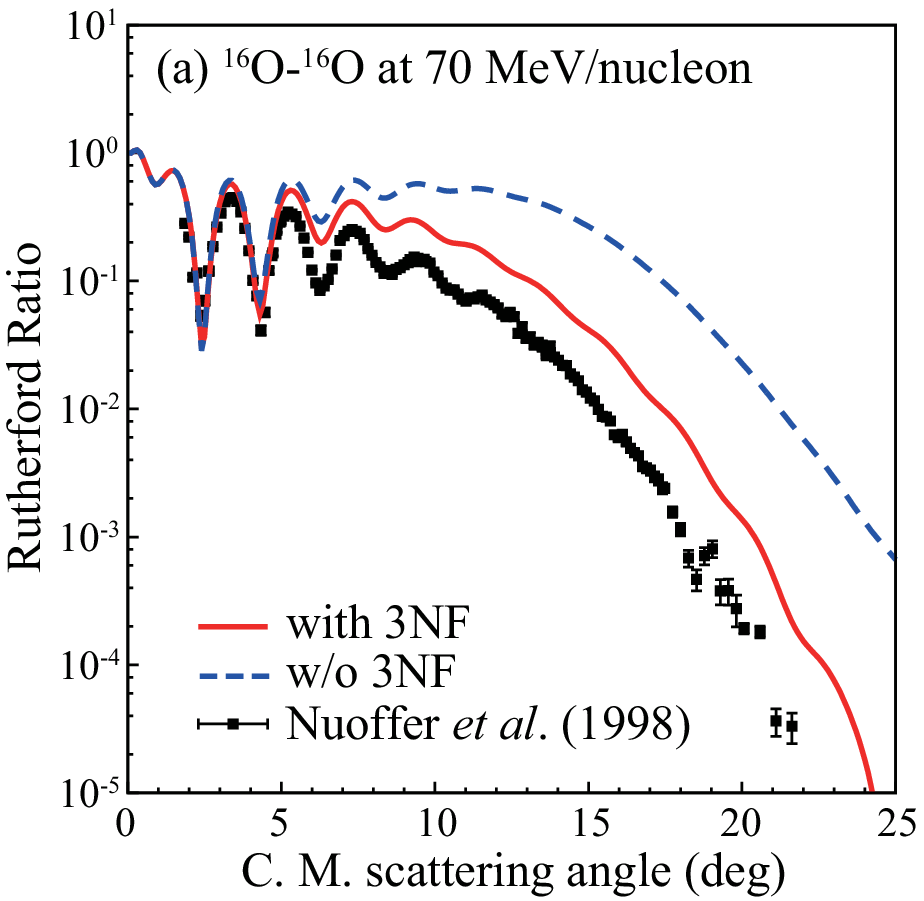}
 \includegraphics[width=0.335\textwidth,clip]{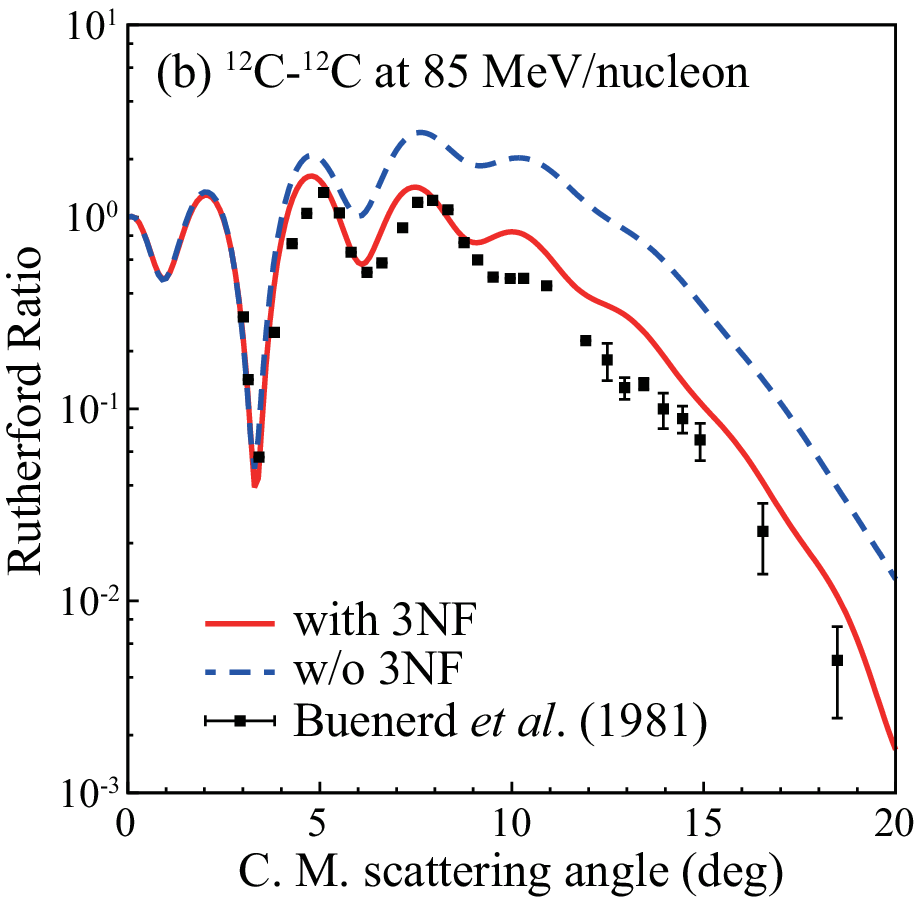}
 \caption{(Color online)
The differential cross sections of
(a) $^{16}$O-$^{16}$O elastic scattering at 70~MeV/nucleon
and (b) $^{12}$C-$^{12}$C elastic scattering at 85~MeV/nucleon
as a function of the scattering angle in the center of mass system.
The solid (dashed) line corresponds to the result of
Melbourne interaction with (without) 3NF corrections.
The experimental data are taken from Refs.~\cite{Nuo98,Bue81}.
}
 \label{fig3}
\end{center}
\end{figure}

Figure~\ref{fig3} shows differential cross sections for 
(a) $^{16}$O-$^{16}$O elastic scattering at 70~MeV/nucleon
and (b) $^{12}$C-$^{12}$C elastic scattering at 85~MeV/nucleon
as a function of scattering angle $\theta$ in the center of mass system.
The solid (dashed) curve corresponds to the result of
Melbourne interaction with (without) chiral-3NF corrections.
Chiral 3NF reduces the cross sections 
in middle and large angles ($\theta > 5^\circ$) and hence 
improves the agreement with the experimental data~\cite{Nuo98,Bue81}.
Switching off chiral-3NF effects for either the real or the imaginary part 
of $U(R)$, we find that 
the interference between the two effects 
is important at the middle and large angles.

%----------------------
% Figure 3
%----------------------
%----------------------

For the $^{16}$O-$^{16}$O scattering, the result of chiral 2NF+3NF slightly 
overestimates the experimental data. 
If the real part of the potential is reduced by $10\%$ and the imaginary 
part is enhanced by $10\%$, the result can reproduce the data. 
Since the correction is rather small, as an origin of the difference we can 
consider higher-order corrections 
such as off-shell corrections due to chiral 3NF and corrections for 
the mean-field approximation. 
For the $^{12}$C-$^{12}$C scattering, 
the diffraction pattern is slightly shifted 
forward in the result of chiral 2NF+3NF compared with the data. 
The difference may come from projectile and/or target excitations, 
since the previous work of Ref.~\cite{Fur13} shows
that the excitation effects slightly shift the diffraction pattern backward. 
The higher-order corrections mentioned above may be masked 
by the excitation effects. 
Further analysis along this line is quite interesting as a future work.

Total reaction cross sections $\sigma_{\rm R}$ are mainly determined 
from the imaginary part of $U(R)$. We then check whether 
the present potential with 3NF corrections reproduces measured 
$\sigma_{\rm R}$. 
Table~\ref{tab1} shows $\sigma_{\rm R}$
for $^{16}{\rm O}$-$^{16}{\rm O}$ scattering at 70 MeV/nucleon
and $^{12}{\rm C}$-$^{12}{\rm C}$ scattering at 85 MeV/nucleon.
The experimental data are available for 
the $^{12}{\rm C}$-$^{12}{\rm C}$ scattering, but not for 
the $^{16}{\rm O}$-$^{16}{\rm O}$ scattering. 
Chiral 3NF enhances $\sigma_{\rm R}$ only by a few percent, 
since the effects appear only at small $R$ 
where $U(R)$ already has strong absorption.
The calculated result slightly overestimates 
the experimental data~\cite{Tak09}, but it is not serious at all.
The Melbourne $g$-matrix interaction with 3NF corrections is thus reliable 
also for $\sigma_{\rm R}$.

%----------------------
% Table 1
%----------------------
\begin{table}[htbp]
\caption{
Total reaction cross sections for nucleus-nucleus scattering
around 70 MeV/nucleon (in the unit of mb).
The experimental data is taken from Ref.~\cite{Tak09}.
}
\begin{center}
\begin{tabular}{ccccc} \hline\hline
 system & & with 3NF & w/o 3NF & Exp.~\cite{Tak09} \\ \hline
 $^{16}$O-$^{16}$O@70MeV/nucleon & & $1423$ & $1401$ & -- \\
 $^{12}$C-$^{12}$C@85MeV/nucleon & & $1025$ & $989$ & $998 \pm 13$ \\ \hline\hline
\end{tabular}
\label{tab1}
\end{center}
\end{table}
%----------------------

%Summary
%\section{Summary}
{\it Summary.} 
We have made a qualitative discussion about the effects of chiral NNLO 3NF 
on nucleus-nucleus elastic scattering, using the simple prescription of 
Eqs.~\eqref{gst} and \eqref{fkf}. 
In this prescription, on-shell corrections due to chiral 3NF are introduced 
to the local Melbourne $g$-matrix interaction. 
It is to be stressed that we have not introduced any ad hoc phenomenological adjustment.

For nuclear matter, chiral 3NF makes 
the single-particle potentials less attractive 
for the $^{1}$E channel and more absorptive for the triplet channels. 
The effects mainly comes from the 2$\pi$ exchange process. 
For the triplet channels, the diagram enhances tensor correlations and hence 
the absorption.
These 3NF corrections are incorporated in the folding potential for 
nucleus-nucleus scattering by modifying the local Melbourne $g$-matrix 
interaction by the multiplicative factor prescribed as Eqs.~\eqref{gst} and \eqref{fkf}. 
The corrections make the folding potential less attractive and more absorptive. 
Since the folding potential is mainly determined from the $^{1}$E- and 
$^{3}$E-channel components, we can conclude that 
the repulsive correction to the folding potential 
comes from the suppression of transitions to $\Delta$ resonance 
and the absorptive correction is originated in the enhancement of 
tensor correlations. 
The two effects reduce differential elastic cross sections 
for both $^{16}$O-$^{16}$O scattering at 70 MeV/nucleon and
$^{12}$C-$^{12}$C scattering at 85 MeV/nucleon.
Eventually, chiral 3NF improves the agreement with the experimental data.

It is instructive to note the qualitative difference of
the present result from other similar studies in the literature.
In the previous works of Refs.~\cite{Fur08,Fur09a,Fur09b,Raf13}, 
the phenomenological 3NFs make the folding potential $U(R)$ 
less attractive and less absorptive. The present result is consistent with 
the previous ones qualitatively for the real part of $U(R)$, 
although the repulsive effect is small in the present work but large 
in the previous works. 
For the imaginary part, meanwhile, chiral 3NF makes $U(R)$ 
more absorptive, whereas the phenomenological 3NFs do $U(R)$ 
less absorptive. 
The difference comes from the fact that chiral 3NF enhances tensor 
correlations but the phenomenological 3NFs do not.

The on-shell corrections due to chiral 3NF surely improve the agreement 
with the experimental data on nucleus-nucleus elastic scattering, 
but the agreement is not perfect. 
This may come from higher-order corrections such as off-shell corrections
due to chiral 3NF and corrections for the mean-field approximation. 
Further analysis along this line toward deeper microscopic understanding
of nucleus-nucleus scattering is quite interesting as a future work.

%Acknowledgement
\section*{Acknowledgements}
This work is supported in part by
by Grant-in-Aid for Scientific Research
(Nos. 244137, 25400266, and 26400278)
from Japan Society for the Promotion of Science (JSPS).

%%--------------------------------------------------------------------%%
%%                           References                               %%
%%--------------------------------------------------------------------%%


\begin{thebibliography}{00}

%- Spin-Orbit Coupling in Heavy Nuclei
\bibitem{Fuj57}
J. Fujita and H. Miyazawa,
Prog. Theor. Phys. {\bf 17}, 366 (1957).

%- Evolution of Nuclear Spectra with Nuclear Forces
\bibitem{Wir02}
R. B. Wiringa and S. C. Pieper,
Phys. Rev. Lett. {\bf 89}, 182501 (2002).

%- Equation of state for dense nucleon matter
\bibitem{Wir88}
R. B. Wiringa, V. Fiks, and A. Fabrocini,
Phys. Rev. C {\bf 38}, 1010 (1988).

%- Two-Body Correlations in Nuclear Systems
\bibitem{Mut00}
H. M\"{u}ther and A. Polls,
Prog. Part. Nucl. Phys. {\bf 45}, 243 (2000).

%- Satulation of Nuclear Matter and Short-Range Correlations
\bibitem{Dew03}
Y. Dewulf, W. H. Dickhoff, D. Van Neck, E. R. Stoddard, and M. Waroquier,
Phys. Rev. Lett. {\bf 90}, 152501 (2003).

%- Is nuclear matter perturbative with low-momentum interactions?
\bibitem{Bog05}
S. K. Bogner, A. Schwenk, R. J. Furnstahl, and A. Nogga,
Nucl. Phys. A {\bf 763}, 59 (2005).

%- The two-nucleon system at next-to-next-to-next-to-leading order
\bibitem{Epe05}
E. Epelbaum, W. Gl\"{o}ckle, and Ulf-G. Mei\ss ner,
Nucl. Phys. A{\bf 747}, 362 (2005).

%- Modern theory of nuclear forces
\bibitem{Epe09}
E. Epelbaum, H.-W. Hammer, and Ulf-G. Mei{\ss}ner,
Rev. Mod. Phys. {\bf 81}, 1773 (2009).

%- Improved nuclear matter calculations from chiral low-momentum intearctions
\bibitem{Heb11}
K. Hebeler, S. K. Bogner, R. J. Furnstahl, A. Nogga, and A. Schwenk, 
Phys. Rev. C {\bf 83}, 031301(R) (2011).

%- Chiral effective field theory and nuclear forces
\bibitem{mac11}
R. Machleidt and D. R. Entem,
Phys. Rep. {\bf 503}, 1 (2011). 

%- Three-nucleon forces from chiral effective field theory
\bibitem{Epe02}
E. Epelbaum, A. Nogga, W. Gl\"{o}ckle, H. Kamada,
Ulf-G. Mei{\ss}ner, and H. Wita{\l}a,
Phys. Rev. C {\bf 66}, 064001 (2002).

%- Spectra and binding energy predictions of chiral interactions for 7Li
\bibitem{Nog06}
A. Nogga, P. Navr$\acute{\rm a}$til, B. R. Barrett, and J. P. Vary,
Phys. Rev. C {\bf 73}, 064002 (2006).

%- Structure of A=10-13 Nuclei with Two- Plus Three-Nucleon Interactions
%   from Chiral Effective Field Theory
\bibitem{Nav07}
P. Navr$\acute{\rm a}$til, V. G. Gueorguiev, J. P. Vary, W. E. Ormand, and A. Nogga,
Phys. Rev. Lett. {\bf 99}, 042501 (2007).

%- Triton with long-range chiral N3LO three-nucleon forces
\bibitem{Ski11}
R. Skibi$\acute{\rm n}$ski \textit{et al}.,
Phys. Rev. C {\bf 84}, 054005 (2011).

%- Dirac-Brueckner-Hartree-Fock versus chiral effective field theory
\bibitem{Sam12}
F. Sammarruca, B. Chen, L. Coraggio, N. Itaco, and R. Machleidt,
Phys. Rev. C {\bf 86}, 054317 (2012).

%- Strength of reduced two-body spin-orbit interaction from
%   a chiral three-nucleon force
\bibitem{Koh12}
M. Kohno,
Phys. Rev. C {\bf 86}, 061301(R) (2012).

%- Symmetric nuclear matter with chiral three-nucleon forces
%   in the self-consistent Green's functions approach
\bibitem{Car13}
A. Carbone, A. Polls, and A. Rios,
Phys. Rev. C {\bf 88}, 044302 (2013).

%- Nuclear and neutron matter G-matrix calculations
%   with a chiral effective field theory potential
%    including effects of three-nucleon interaction
\bibitem{Koh13}
M. Kohno,
Phys. Rev. C {\bf 88}, 064005 (2013).

% JLM
\bibitem{Jeu77}
J. P. Jeukenne, A. Lejeune and C. Mahaux, Phys. Rev. C{\bf 16}, 80 (1977).

% M3Y
\bibitem{Ber77}
G. Bertsch, J. Borysowicz, H. McManus, and W.G. Love,
Nucl. Phys. A{\bf 284}, 399(1977).

\bibitem{Bri77}
F.A. Brieva and J.R. Rook, Nucl. Phys. A{\bf 291}, 299 (1977);
ibid. 291, 317 (1977); ibid. 297, 206 (1978).

% CEG
\bibitem{Yam83}
N. Yamaguchi, S. Nagata and T. Matsuda,
Prog. Theor. Phys. {\bf 70}, 459 (1983);
N. Yamaguchi, S. Nagata and J. Michiyama,
Prog. Theor. Phys. {\bf 76}, 1289 (1986).

%- Melbourne g-matrix
\bibitem{Amo00}
K. Amos, P. J. Dortmans, H. V. von Geramb, S. Karataglidis,
and J. Raynal, Adv. Nucl. Phys. {\bf 25}, 275 (2000).

%- Generalized folding model for elastic and inelastic nucleus-nucleus scattering
%   using realistic density dependent nucleon-nucleon interaction
\bibitem{Kho00}
D. T. Khoa and G. R. Satchler,
Nucl. Phys. A {\bf 668}, 3 (2000).

% ESC
%- Extended-soft-core baryon-baryon model.
%   I. Nucleon-nucleon scattering with the ESC04 interaction
\bibitem{Rij06a}
Th. A. Rijken,
Phys. Rev. C {\bf 73}, 044007 (2006).

% ESC
%- Extended-soft-core baryon-baryon model.
%   II. Hyperon-nucleon interaction
\bibitem{Rij06b}
Th. A. Rijken and Y. Yamamoto,
Phys. Rev. C {\bf 73}, 044008 (2006).

% CEG07
%- New complex G-matrix interactions derived from two- and three-body forces
%   and application to proton-nucleus elastic scattering
\bibitem{Fur08}
T. Furumoto, Y. Sakuragi, and Y. Yamamoto,
Phys. Rev. C {\bf 78}, 044610 (2008).

%- Three-body-force effect on nucleus-nucleus elastic scattering
\bibitem{Fur09a}
T. Furumoto, Y. Sakuragi, and Y. Yamamoto,
Phys. Rev. C {\bf 79}, 011601(R) (2009).

%- Effect of repulsive and attractive three-body forces
%   on nucleus-nucleus elastic scattering
\bibitem{Fur09b}
T. Furumoto, Y. Sakuragi, and Y. Yamamoto,
Phys. Rev. C {\bf 80}, 044614 (2009).

%- Equation of state and the nucleon optical potential
%   with three-body force
\bibitem{Raf13}
S. Rafi, M. Sharma, D. Pachouri, W. Haider, and Y. K. Gambhir,
Phys. Rev. C {\bf 87}, 014003 (2013).

% Urbana IX
%- Quantum Monte Carlo calculations of nuclei with A<7
\bibitem{Pud97}
B. S. Pudliner, V. R. Pandharipande, J. Carlson, S. C. Pieper, and R. B. Wiringa,
Phys. Rev. C {\bf 56}, 1720 (1997).

% TNI
%- Hot and cold, nuclear and neutron matter
\bibitem{Fri81}
B. Friedman and V. R. Pandharipande,
Nucl. Phys. A {\bf 361}, 502 (1981).

% TNI
%- Variational calculations of realistic models of nuclear matter
\bibitem{Lag81}
I. E. Lagaris and V. R. Pandharipande,
Nucl. Phys. A {\bf 359}, 349 (1981).

% AV18
%- Acculate nucleon-nucleon potential
%       with charge-independence breaking
\bibitem{Wir95}
R. B. Wiringa, V. G. J. Stoks, and R. Schiavilla,
Phys. Rev. C {\bf 51}, 38 (1995).

%- Nucleon-nucleon correlation and two-pion-exchange three-body force in nuclear matter
\bibitem{Loi71}
B. A. Loiseau, Y. Nogami, and C. K. Ross,
Nucl. Phys. A {\bf 165}, 601 (1971). 

%- Effects of three-body force in nuclear matter
\bibitem{Kas74}
T. Kasahara, Y. Akaishi, and H. Tanaka,
Prog. Theor. Phys. Suppl. {\bf 56}, 96 (1974).

%- Density-dependent effective nucleon-nucleon interaction
%   from chiral three-nucleon forces
\bibitem{Hol10}
J. W. Holt, N. Kaiser, and W. Weise,
Phys. Rev. C {\bf 81}, 024002 (2010).

%- The Bonn meson-exchange model for nucleon-nucleon interaction
\bibitem{Mac87}
R. Machleidt, K. Holinde, and Ch. Elster,
Phys. Rep. {\bf 149}, 1 (1987).

%- The Brieva-Rook Localization of
%   the Microscopic Nucleon-Nucleus Potential
\bibitem{Min10}
K. Minomo, K. Ogata, M. Kohno, Y. R. Shimizu, and M. Yahiro,
J. Phys. G {\bf 37}, 085011 (2010)

%- Mass-number and isotope dependence of local microscopic optical potentials
%   for polarized proton scattering
\bibitem{Toy13}
M. Toyokawa, K. Minomo, and M. Yahiro,
Phys. Rev. C {\bf 88}, 054602 (2013).

%- DFM-standard-form
\bibitem{Sin75}
B. Sinha,
Phys. Rep. {\bf 20}, 1 (1975).

\bibitem{Sin79}
B. Sinha and S. A. Moszkowski,
Phys. Lett. B {\bf 81}, 289 (1979).

%- Folding model potentials from realistic interactions for heavy-ion scattering
\bibitem{Sat79}
G. R. Satchler and W. G. Love,
Phys. Rep. {\bf 55}, 183 (1979).

% phenomenological densities
%- Nuclear charge-density-distribution parameters
%   from elastic electron scattering
\bibitem{Vri87}
H. de Vries, C. W. de Jager, and C. de Vries,
At. Data Nucl. Data Tables {\bf 36}, 495 (1987).

% proton-radius-effect
%- Folding of proton size in nuclear structure calculations
\bibitem{Sin78}
R. P. Singhal, M. W. S. Macauley, and P. K. A. de Witt Huberts,
Nucl. Instr. and Meth. {\bf 148}, 113 (1978).

%- The equation of state for cold nuclear matter
%   as seen in nucleus-nucleus scattering
\bibitem{Nuo98}
F. Nuoffer \textit{et al}.,
Nuovo Cimento A {\bf 111}, 971 (1998).

%- Elastic and inelastic scattering of 1.03 GeV 12C projectiles
\bibitem{Bue81}
M. Buenerd \textit{et al}.,
Phys. Lett. B {\bf 102}, 242 (1981).

%- Channel coupling effect and important role of imaginary part of
%   coupling potential for high-energy heavy-ion scattering
\bibitem{Fur13}
T. Furumoto and Y. Sakuragi,
Phys. Rev. C {\bf 87}, 014618 (2013).

%- Reaction cross sections at intermediate energies and Fermi-motion effect
\bibitem{Tak09}
M. Takechi \textit{et al}.,
Phys. Rev. C {\bf 79}, 061601(R) (2009).

\end{thebibliography}
\end{document}